\documentclass[review]{elsarticle}

\usepackage{lineno,hyperref}
\modulolinenumbers[5]
\usepackage[margin=1in]{geometry}

\usepackage{graphicx}
\usepackage{listings}
\usepackage{makecell}
\usepackage[utf8]{inputenc}
\usepackage[T1]{fontenc}
\usepackage{amsmath}
\usepackage{tikz}
\usetikzlibrary{arrows.meta}

\usepackage{booktabs}
\usepackage{algorithm}

\usepackage{algpseudocode}

\usepackage{tcolorbox}
\usepackage{listings}
\usepackage{caption}

\usepackage{sectsty}
\sectionfont{\fontsize{10}{10}\selectfont}

\usepackage{listings}
\usepackage{xcolor}

\usepackage[numbers,sort&compress]{natbib}
\biboptions{numbers,sort&compress}

\lstdefinestyle{promptstyle}{
    basicstyle=\ttfamily\small,
    breaklines=true,
    frame=single,
    columns=fullflexible,
    keepspaces=true,
    showstringspaces=false
}

\usepackage{amssymb}
\usepackage{lscape}
\usepackage{tabularx}

\begin{document}

\begin{frontmatter}



\title{ACE: An AI-Orchestrated Intrusion Detection and Compliance Mapping Engine }



\author[1]{Abdoulie Ceesay}
\ead{acees@uis.edu}

\author[1]{Rishikesh Sahay}
\ead{rsaha@uis.edu}

\author[1]{Md Rasel Al Mamun}
\ead{mmamu@uis.edu}

\author[1]{Bell Raj Eapen}
\ead{bpunn@uis.edu}


\address[1]{Department of Management Information Systems, University of Illinois, Springfield, USA}



\begin{abstract}

This study proposes the Automated Compliance Engine (ACE), an AI-orchestrated framework that links intrusion detection, regulatory control mapping, and automated countermeasure deployment for security operations environments.
ACE integrates Suricata IDS, a LangGraph five-stage workflow, retrieval-augmented generation, NIST SP 800-53 Rev. 5 control retrieval, large language model-based control ranking, and SSH-based Suricata rule deployment. 
The framework was evaluated in a two-virtual-machine laboratory using eight simulated attack scenarios and three legitimate traffic tests.
ACE processed all eight attack alerts, achieved 100\% alert detection, 0\% false positives across legitimate traffic tests, generated deployable AUTO rules for 37.5\% of alerts, verified 100\% deployment of generated rules, and achieved an average NIST control-mapping confidence score of 0.87. 
The DeepEval assessment produced 1.0 scores for answer relevancy and faithfulness, while the hallucination score reflected known LLM-as-judge calibration limitations.
The evaluation was conducted in a controlled environment with a limited attack set.
Further validation is required using enterprise-scale telemetry, multiple IDS sensors, and an analyst-in-the-loop review.
The results demonstrate how AI-assisted security automation can reduce analyst effort in translating IDS alerts into compliance-relevant evidence and defensive actions.
ACE contributes an integrated detection-to-compliance-to-response workflow combining IDS telemetry, RAG-based control mapping, and automated rule deployment.

\end{abstract}




\end{frontmatter}



\section{Introduction}
\label{sec:intro}

Security operations centers (SOCs) increasingly operate under two simultaneous pressures. First, modern networks generate high volumes of intrusion detection system (IDS), endpoint, and application telemetry that must be triaged rapidly. Second, organizations must document how detected security events affect regulatory and control obligations. These two functions are often treated separately. IDS tools such as Suricata can identify suspicious activity, while governance, risk and compliance processes require analysts to translate technical alerts into control-level evidence. The result is a persistent semantic and operational gap between alert detection and compliance documentation.

The Automated Compliance Engine (ACE) addresses this gap by integrating threat detection, compliance mapping, and countermeasure deployment into a single auditable workflow. ACE uses Suricata IDS 8.0.3 to collect network security alerts, a LangGraph-based orchestration layer to manage a five-stage workflow, a retrieval-augmented generation (RAG) architecture to map interpreted threats to NIST SP 800-53 Rev. 5 security controls, and an SSH-based deployment process to push generated Suricata rules to a target system. The system was implemented in a two-state virtual-machine lab consisting of a Kali Linux orchestrator and a Windows IDS host.

The research problem addressed by this study is therefore practical and architectural: how can a system automatically interpret a low-level IDS alert, retrieve relevant security controls, score the relevance of those controls, generate targeted countermeasure rules from observable network indicators, and deploy those rules without producing duplicate or unnecessary changes across repeated runs? This question is important because manual translation of alerts into compliance evidence can be slow, inconsistent, and difficult to reproduce. It also becomes harder when alerts contain heterogeneous metadata from HTTP, DNS, TLS, or transport-layer events.

Although compliance automation is often treated as an administrative problem, it is also a security problem. Alerts contain technical evidence that must be interpreted, contextualized and preserved. If this evidence remains isolated within IDS logs, governance teams must reconstruct the connection manually after an incident. ACE treats compliance mapping as part of the detection workflow itself. In this framing, the security event is not complete until it has been associated with relevant controls and a documented response path.

The emphasis on automated deployment also reflects a shift from passive compliance reporting to active security control reinforcement. A system that can only describe compliance relevance may improve reporting, but it does not directly change the monitored environment. By contrast, ACE can generate narrowly scoped Suricata rules when alert metadata contains actionable indicators. This creates a feedback loop in which detection produces both documentation and a candidate technical control. The framework therefore joins three activities that are frequently separated in practice: monitoring, governance, and defensive engineering.

ACE is positioned as a defensive AI automation framework rather than a general chatbot or advisory tool. The workflow limits model autonomy by grounding control mapping in a NIST vector database, routing execution through a deterministic supervisor node, and validating rule generation through a structured rule registry. These design choices address well-known concerns about large language model (LLM) hallucination, non-determinism, and unsafe autonomous action in security settings.

This study makes three contributions. First, it introduces an integrated architecture that connects IDS alert interpretation, NIST SP 800-53 Rev. 5 control mapping, and Suricata rule deployment. Second, it demonstrates how RAG can reduce the semantic gap between raw IDS events and high-level compliance controls by retrieving relevant control text before confidence scoring. Third, it evaluates the framework using eight simulated attack scenarios and three legitimate traffic tests, reporting detection accuracy, false-positive behavior, rule deployment success, NIST mapping confidence, and LLM output quality.

The remainder of the paper is organized as follows. Section 2 reviews related work on IDS, compliance mapping, RAG and agentic AI in cybersecurity. Section 3 presents the research gap and the conceptual framing of ACE. Section 4 describes the system architecture. Section 5 explains the methodology and experimental design. Section 6 reports the evaluation results. Section 7 discusses implications for SOC operations and compliance. Section 8 presents limitations and future work. Section 9 concludes the paper.

\section{Related Work}
\label{sec:related_works}

IDS platforms remain foundational to network security monitoring. Suricata is an open-source IDS, IPS, and network security monitoring engine maintained by the Open Information Security Foundation. It produces structured event logs, including eve.json records that capture source and destination addresses, protocol fields, alert signatures, SIDs, and application-layer metadata. These structured records make Suricata a useful sensor for automated security workflows, but Suricata by itself does not interpret the compliance meaning of an alert or produce audit-ready control mappings.

Security Information and Event Management (SIEM) systems aggregate alerts from IDS, endpoint, and infrastructure sources. SIEM platforms can correlate and enrich events, but many organizations still rely on analysts to connect technical events to governance frameworks. This distinction matters because regulatory frameworks are written in high-level language, whereas IDS alerts are low-level observations. Mapping an event such as a SYN scan, SQL injection attempt, or DNS abuse event to NIST SP 800-53 Rev. 5 requires semantic interpretation, not only pattern matching.

NIST SP 800-53 Rev. 5 provides a broad catalog of security and privacy controls organized into control families such as Audit and Accountability, Incident Response, Risk Assessment, System and Communications Protection, and System and Information Integrity. It is widely used in federal and regulated environments and is also referenced by private-sector organizations as a security-control baseline. However, the control text is not written in the same language as IDS signatures. This creates a semantic gap that is well suited to retrieval and language-model-based interpretation.

Recent work has explored the use of LLMs and RAG for compliance and security reasoning. 
RAG improves factual grounding by retrieving relevant documents at inference time and providing them as context for the model~\cite{ragnlp}.
Das et al.~\cite{multiagentrag} shows that RAG and multi-agent architectures can support automated compliance verification by reducing reliance on the model's parametric knowledge alone. 
Masoudifard et al.~\cite{promptsforcompliance} further demonstrate that graph-enhanced RAG can improve compliance-related retrieval by representing relationships among higher-level requirements.
These studies support the assumption that compliance mapping benefits from retrieval-based reasoning.

LLM-enabled security systems also introduce risks. 
Security tasks require precision, reproducibility, and traceability. The authors in~\cite{benchmarkingllm} report that LLM performance in threat modeling varies across models and task dimensions, suggesting that unrestricted model output should not be treated as authoritative. 
Peterson (2025) shows that even structured outputs can contain silent omissions or mistranslations when LLMs normalize indicators into structured formats. These findings support the use of deterministic validation layers around LLM components, especially when outputs influence security controls or compliance records.

Agentic AI frameworks are increasingly used to coordinate multi-step security workflows.
LangGraph supports graph-based orchestration with stateful nodes and deterministic or model-driven routing.
The authors in~\cite{redteamllm} demonstrate how agentic workflows can coordinate reconnaissance, enumeration, and offensive security tasks. 
ARGOS applies LLM-based agentic rule generation to cloud anomaly detection and shows the value of separating generation, review, and repair activities~\cite{argos}.
Although these works focus on different security domains, they indicate that stateful orchestration is useful when an AI workflow must transform observations into actions.

Governance of agentic AI is especially important in security operations. 
The framework TRiSM provides a Trust, Risk, and Security Management framework for LLM-based agentic multi-agent systems and identify risks such as prompt injection, persistent memory compromise, autonomy abuse, orchestration exploits, and agent collusion~\cite{trism}. 
Evani.et.al.~\cite{evani} similarly identifies goal-inversion attacks, adversarial feedback loops, and decision-cascade failures as risks in autonomous decision-making systems.
ACE addresses these concerns through a deterministic supervisor node, an append-only SID registry, constrained rule templates, and a workflow that keeps LLM outputs separate from final deployment logic.

Several commercial platforms provide partial automation in this space through playbooks, SOAR actions, and compliance dashboards. However, playbooks are often manually authored, and compliance dashboards commonly rely on asset or configuration evidence rather than semantic mapping from live alerts. ACE differs by using an AI-guided interpretation layer between raw network events and control documentation. The purpose is not to replace SOAR or SIEM platforms but to show how an AI compliance-mapping component could be integrated into them.

Another related area is explainable AI for cybersecurity. Security analysts need to know why a recommendation was produced, not only what the recommendation is. 
ACE supports explainability through structured state transitions and by preserving retrieved control text alongside rankings. This is different from a black-box classifier that outputs a label without traceable evidence. The explanation is distributed across the workflow: the alert explains the observation, the threat description explains the interpretation, retrieved controls explain the compliance basis and rule templates explain the response action.

The literature on human-in-the-loop AI is also important. In high-risk domains, full autonomy can be inappropriate even when model outputs appear accurate. ACE currently executes deployment automatically in a controlled laboratory, but its architecture can support human review by placing approval between Generate and Deploy. This would allow analysts to review the interpreted threat, retrieved controls, and proposed rules before they are installed. Such a design would preserve the speed of automated interpretation while maintaining accountability for operational change.

Compliance mapping has historically depended on expert judgment because controls are intentionally broad. For example, a NIST control may require monitoring, detection, incident handling, or boundary protection without prescribing a single alert signature. This flexibility helps organizations adapt controls to different environments, but it also makes automated mapping difficult. A keyword search for terms such as "scan," "injection," or "malware" may retrieve some relevant controls, but it may miss semantic relationships between attacker behavior and control intent. RAG provides a middle ground by retrieving candidate controls based on semantic similarity and allowing the ranking step to evaluate contextual relevance.

Domain adaptation and machine learning robustness are also relevant to the ACE motivation. SIEM and IDS environments often experience distribution shifts as network services, users, devices, and adversary techniques change. Although ACE does not implement a source-free domain adaptation model, prior work in that area highlights the broader challenge of deploying models across changing operational domains. The lesson for ACE is that security automation must avoid assuming a static environment. This is one reason the system retrieves control text dynamically, maintains a registry of processed alerts, and separates alert interpretation from rule deployment.

Prior research therefore supports the individual components of ACE: IDS telemetry, RAG-based grounding, LLM-assisted interpretation, graph-based orchestration, and rule-generation automation. However, these components are often studied separately. Existing IDS tools detect alerts but do not map them to compliance controls; compliance-mapping tools often lack live alert ingestion and rule deployment; offensive agentic frameworks demonstrate automation but are not designed for defensive compliance workflows. ACE combines these functions into one pipeline and evaluates their interaction in a live laboratory environment.

\section{Research gap and contribution framing}

The central gap addressed in this paper is the absence of an integrated, auditable workflow that connects network intrusion detection with compliance mapping and defensive rule deployment. Traditional IDS tools detect events but do not produce compliance evidence. SIEM workflows centralize logs but often require manual analyst interpretation. RAG-based compliance systems improve mapping accuracy but are commonly disconnected from live IDS telemetry and response actions. Agentic cybersecurity systems can automate multi-step tasks, but many focus on offensive reconnaissance or isolated detection-rule generation rather than detection-to-compliance-to-response integration.

Table~\ref{tab:comparison} compares ACE with traditional IDS tools and representative LLM-based approaches across the capabilities relevant to this gap.

\begin{table}[htbp]
\centering
\caption{Comparison with existing approaches}
\label{tab:comparison}
\scriptsize
\setlength{\tabcolsep}{2pt}
\begin{tabular}{>{\raggedright\arraybackslash}p{0.16\textwidth} >{\raggedright\arraybackslash}p{0.125\textwidth} >{\raggedright\arraybackslash}p{0.125\textwidth} >{\raggedright\arraybackslash}p{0.125\textwidth} >{\raggedright\arraybackslash}p{0.125\textwidth} >{\raggedright\arraybackslash}p{0.125\textwidth} >{\raggedright\arraybackslash}p{0.125\textwidth}}
\toprule
\textbf{Feature} & \textbf{ACE} & \textbf{Trad. IDS} & \textbf{Das et al.} & \textbf{\makecell[l]{Masoudifard\\et al.}} & \textbf{RedTeamLLM} & \textbf{Sekercan} \\
\midrule
Real-time alert detection & Yes & Yes & No & No & N/A (offensive) & No (offline) \\
Automated threat interpretation & Yes (LLM) & No & Yes (LLM) & Yes (LLM) & Yes (LLM) & Yes (LLM) \\
Compliance/framework mapping & Yes (RAG+LLM) & No & Yes (RAG+LLM) & Yes (Graph-RAG) & No & MITRE ATT\&CK \\
Advanced prompt reasoning & No & No & No & CoT+ToT & ReAct & CoT \\
Autonomous rule generation & Yes & No & No & No & No & No \\
Automated rule deployment & Yes (SSH) & No & No & No & No & No \\
Cross-run deduplication & Yes (SID) & No & No & No & No & No \\
TRiSM/governance framework & Partial & No & Partial & No & Partial & No \\
Human-in-the-loop validation & No & No & Yes & No & Yes & No \\
\bottomrule
\end{tabular}
\end{table}

As Table~\ref{tab:comparison} shows, existing approaches cover individual capabilities such as threat interpretation or compliance mapping, but none combines live alert detection, compliance mapping, rule generation, automated deployment and cross-run deduplication in one workflow. ACE does not provide human-in-the-loop validation or advanced prompt reasoning in its current form; these are addressed as future work in Section 8.

ACE is framed as a security information systems artifact. It is not intended to replace analysts or enterprise governance processes. Instead, it automates repetitive translation steps that occur after an IDS alert is observed: interpreting the alert, retrieving relevant controls, ranking the controls, proposing a countermeasure and preserving an audit trail. The design objective is to reduce manual friction while maintaining determinism where operational safety requires it.

Three design principles guide the artifact. The first is grounded autonomy: LLMs may interpret and rank information, but their outputs should be grounded in retrieved documents and constrained by deterministic templates. The second is operational idempotence: repeated pipeline runs should not create duplicate rules or repeated deployments for the same alert. The third is auditability: each major decision should be traceable to an input alert, retrieved control, or registry entry. These principles reflect the requirements of both security operations and compliance evidence generation.

The artifact also reflects a boundary between recommendation and execution. The Interpret and Rank nodes produce semantic outputs that are probabilistic by nature. The Generate and Deploy nodes produce operational changes and therefore rely on deterministic extraction, templates, and registry checks. This separation is important because it prevents free-form model text from becoming executable security policy without validation.

The following research questions guide the evaluation. RQ1 asks whether ACE can ingest and process Suricata IDS alerts and map them to relevant NIST SP 800-53 Rev. 5 controls. RQ2 asks whether ACE can generate and deploy valid Suricata countermeasure rules from available alert metadata. RQ3 asks what limitations arise when LLM and RAG components are used for compliance mapping and automated security response. These questions align with the journal audience because they concern information security system design, operational feasibility and the governance implications of AI-enabled security automation.

The expected contribution is empirical rather than purely conceptual. The system is implemented and evaluated in a controlled two-state VM environment with real Suricata alerts, generated rules, and deployment verification. The results should therefore be interpreted as a proof-of-concept demonstration of feasibility, not as evidence that ACE is production-ready in its current form.

\section{ACE system architecture}
\begin{figure*}[!t]
    \centering
    \includegraphics[width=\textwidth]{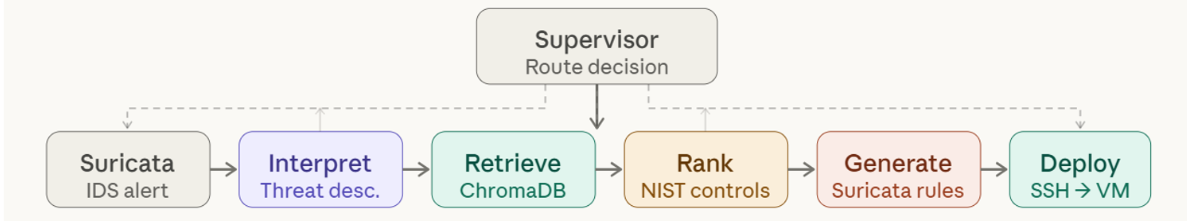}
    \caption{ACE five-stage AI-orchestrated pipeline. The deterministic Supervisor routes each alert through the Interpret, Retrieve, Rank, Generate and Deploy stages.}
    \label{doc:ace-workflow}
\end{figure*}

Figure~\ref{doc:ace-workflow} shows the operational workflow of the five-stage pipeline. An alert generated by the Suricata IDS on the Windows host is exposed to the Kali orchestrator through a shared folder. The orchestrator checks the alert's SID against the rule registry, interprets the alert, retrieves and ranks candidate NIST SP 800-53 Rev. 5 controls, generates constrained Suricata rules, and deploys them back to the IDS host over SSH.

ACE runs across two virtual machines connected through a host-only network. The Kali Linux VM functions as the orchestrator, LLM client, RAG host and attack-simulation platform. The Windows VM functions as the Suricata IDS host and the destination for generated rule deployment. The two systems communicate through SSH for rule deployment and through shared folders for log and proposal-file exchange. This compact architecture was selected to provide a reproducible laboratory environment while preserving the separation between the monitoring host and the orchestration host.

The architecture consists of four functional layers. The sensing layer is Suricata IDS, which generates eve.json events. The interpretation and retrieval layer consists of the LangGraph pipeline, ChromaDB vector database and embedding model. The action layer generates Suricata rules and deploys them to the Windows host. The governance layer consists of the deterministic Supervisor node, the persistent rule registry and the SID-based deduplication logic. 
Table~\ref{tab:components} (Section 5) summarizes the core components.

\begin{table}[htbp]
\centering
\caption{LangGraph pipeline node descriptions}
\label{tab:nodes}
\small
\begin{tabular}{>{\raggedright\arraybackslash}p{0.15\textwidth} >{\raggedright\arraybackslash}p{0.14\textwidth} >{\raggedright\arraybackslash}p{0.50\textwidth} >{\raggedright\arraybackslash}p{0.11\textwidth}}
\toprule
\textbf{Node} & \textbf{Type} & \textbf{Function} & \textbf{LLM call} \\
\midrule
Supervisor & Router & Checks workflow state and routes execution deterministically & No \\
Interpret & Processor & Creates structured threat description from Suricata alert & Yes \\
Retrieve & Processor & Queries ChromaDB for relevant NIST controls & No \\
Rank & Processor & Scores and ranks the top three retrieved controls & Yes \\
Generate & Processor & Extracts indicators and creates Suricata rule proposals & No \\
Deploy & Processor & Writes and deploys rules to the Windows IDS host & No \\
\bottomrule
\end{tabular}
\end{table}

The LangGraph workflow is implemented as a StateGraph with five processing nodes and one deterministic Supervisor. The Supervisor acts as the routing hub and checks state fields to determine the next node. Unlike an LLM-driven orchestrator, it does not decide dynamically based on free-form model output. This design reduces the possibility of decision-cascade errors and makes the execution path predictable.

Table~\ref{tab:nodes} summarizes the nodes.

The Interpret node receives a raw Suricata alert and creates a structured threat description using the alert signature, source and destination IP addresses, protocol, and any available HTTP, TLS, or DNS fields. The output is a plain-language but structured description of the suspected activity, attacker objective, and potential system impact. The Retrieve node embeds the threat description using the all-MiniLM-L6-v2 sentence-transformers model and queries a ChromaDB vector database containing parsed NIST SP 800-53 Rev. 5 controls. The top three controls are returned with metadata such as control ID, family, and control text.

The Rank node uses the threat description and retrieved controls to assign confidence scores between 0 and 1. The output contains the control ID, rank, confidence score, and concise reasoning. The Generate node reexamines the original alert and extracts available indicators using a priority order: HTTP user-agent, HTTP host, HTTP URI, TLS SNI, DNS query name, payload tokens, and finally source IP fallback. The Deploy node writes newly generated rules to a proposed rules file and transfers them to the Windows IDS host over SSH.

SID-based deduplication is central to ACE's operational safety. The first deduplication level checks whether an alert's signature\_id is already present in the rule\_registry.json file. If the SID has already been processed, the alert is skipped, and no LLM calls are made. The second level relies on rule fingerprinting to prevent repeated generation of equivalent rules. This approach reduces unnecessary deployment, limits repeated LLM expense, and provides an append-only audit trail of processed alerts and generated countermeasures.

Security controls around the deployment channel are critical in any production implementation. The laboratory uses SSH between the orchestrator and Windows host, but an enterprise deployment should use least-privilege service accounts, key rotation, command restrictions, logging, and rollback. ACE's design does not require broad administrative autonomy; it requires only the ability to place validated rule files in an approved Suricata rule path and trigger a controlled reload or restart process.

The architecture also separates batch-level and alert-level operations. Interpretation, retrieval, ranking and generation occur per alert, while final deployment can be performed for the accumulated rule set. This reduces the number of SSH sessions and lowers the likelihood of partial deployment. It also allows the system to produce a single proposed\_rules.rules artifact that can be reviewed or archived as part of the run output.

State management is another important architectural feature. Each alert is represented as a state object that accumulates fields as it moves through the graph. The original alert is retained so that later nodes can refer back to authoritative metadata rather than relying on the LLM's paraphrase. This reduces the risk that an interpretation error will be amplified into the rule-generation stage. The threat description is used for retrieval and ranking, but indicator extraction always returns to the original Suricata event.

The rule-generation strategy is intentionally constrained. ACE does not allow the LLM to write arbitrary Suricata syntax. Instead, deterministic templates are used for specific indicator types. The generated rules use a reserved SID range, the message prefix AUTO, and a predefined class type. This approach allows the system to use AI for interpretation and ranking while keeping the final defensive action within a controlled template space.

\section{Experimental Design}

The evaluation used a controlled laboratory experiment designed to test the full pipeline from alert ingestion to compliance mapping and rule deployment. The experiment included eight simulated attack scenarios across four categories: reconnaissance, web application attacks, DNS abuse, and denial-of-service activity. Three legitimate traffic tests were also performed to assess false-positive behavior. The evaluation was designed to measure not only detection but also rule-generation feasibility, deployment verification, NIST mapping quality, and LLM output quality.

Table~\ref{tab:components} summarizes the laboratory components used in the experiments.

\begin{table}[htbp]
\centering
\caption{ACE system component overview}
\label{tab:components}
\small
\begin{tabular}{>{\raggedright\arraybackslash}p{0.22\textwidth} >{\raggedright\arraybackslash}p{0.20\textwidth} >{\raggedright\arraybackslash}p{0.50\textwidth}}
\toprule
\textbf{Component} & \textbf{Host or location} & \textbf{Function} \\
\midrule
Kali Linux & 10.0.2.3 & Pipeline execution, LLM interface, RAG host and attack simulation \\
Windows VM & 10.0.2.15 & Suricata IDS host and target for rule deployment \\
SSH & Kali to Windows & Secure transfer and installation of proposed rules \\
CIFS shared folders & Windows share mounted on Kali & Access to eve.json and proposed rule files \\
ChromaDB & Kali local & Vector store for NIST SP 800-53 Rev. 5 controls \\
rule\_registry.json & Kali local & Persistent SID and rule deduplication registry \\
GPT-4.1 API & Remote API & Threat interpretation and control ranking \\
\bottomrule
\end{tabular}
\end{table}

Table~\ref{tab:attacks} lists the eight attack scenarios.
Reconnaissance was represented by Nmap SYN, FIN and XMAS scans and an ICMP echo test. Web application activity was represented by SQL injection tooling and suspicious HTTP user-agent behavior. DNS abuse was represented by a suspicious domain lookup. Denial-of-service activity was represented by a UDP flood. These scenarios were selected because they generate different metadata profiles in Suricata alerts. Application-layer attacks often include extractable HTTP or DNS indicators, whereas network-layer scans may contain fewer fields suitable for indicator-specific rule generation.

\begin{table}[htbp]

\centering

\caption{Attack scenarios and tool categories}
\label{tab:attacks}

\small

\begin{tabular}{>{\raggedright\arraybackslash}p{0.24\textwidth} >{\raggedright\arraybackslash}p{0.16\textwidth} >{\raggedright\arraybackslash}p{0.18\textwidth} >{\raggedright\arraybackslash}p{0.12\textwidth} >{\raggedright\arraybackslash}p{0.22\textwidth}}

\toprule

\textbf{Attack scenario} & \textbf{Tool} & \textbf{Category} & \textbf{Local SID} & \textbf{Generated indicators} \\

\midrule

Nmap SYN port scan & nmap -sS & Reconnaissance & 1000031 & Source IP fallback \\

Nmap FIN scan & nmap -sF & Reconnaissance & 1000008 & No generated rule \\

Nmap XMAS scan & nmap -sX & Reconnaissance & 1000006 & No generated rule \\

ICMP echo test & ping & Reconnaissance & 1000013 & No generated rule \\

Suspicious DNS query & nslookup malware.com & DNS abuse & 1000032 & No generated rule \\

SQL injection & sqlmap / curl & Web application & 1000034 & HTTP user-agent, host and URI \\

Suspicious HTTP user-agent & sqlmap & Web application & 1000020 & HTTP user-agent and host \\

UDP flood & BoNeSi & Denial of service & 1000035 & No generated rule \\

\bottomrule

\end{tabular}

\end{table}

The RAG knowledge base was created by parsing NIST SP 800-53 Rev. 5 controls into vector-ready documents. During the offline phase, controls were embedded using all-MiniLM-L6-v2, which produces 384-dimensional embeddings, and stored in ChromaDB using an HNSW index. During the online phase, each LLM-generated threat description was embedded and used to retrieve the three most similar controls. The retrieved controls were then ranked by the Rank node. This workflow ensured that control ranking was grounded in retrieved NIST text rather than relying solely on the model's internal knowledge.

The deployment evaluation was based on generated AUTO rules. For alerts with extractable indicators, the Generate node produced Suricata rules and assigned SIDs in the reserved range. These rules were written to proposed\_rules.rules and deployed to the Windows host. Deployment success was verified by checking whether the expected rules appeared in the Windows signatures.rules file after deployment. This metric tests whether the pipeline can complete an end-to-end defensive action, not only produce a local proposal.

False-positive testing used three legitimate activities: a normal HTTP GET request, wget traffic and a normal DNS lookup. The evaluation used a delta snapshot approach rather than clearing eve.json. The set of alert signatures present before legitimate tests was recorded, the legitimate tests were executed, and the post-test set was compared with the pre-test set. New alert signatures appearing only after legitimate traffic would be counted as false positives. This approach preserved prior attack evidence while still measuring new alerts induced by benign activity.

LLM output quality was evaluated with DeepEval metrics for answer relevancy, faithfulness, and hallucination. These metrics were used as supporting evidence rather than as the sole basis for correctness. This distinction is important because LLM-as-judge methods can vary under abstention policies and threshold settings. For that reason, ACE treats DeepEval results as one quality indicator alongside deterministic checks such as SID matching, rule presence, and deployment verification.

The evaluation deliberately preserved negative cases for rule generation. A tempting but misleading evaluation would count only alerts that contain extractable application-layer indicators. ACE instead reports rule generation across all eight alerts, producing a more conservative rate of 37.5\%. This better reflects real SOC conditions, where some alerts can produce specific blocking or detection rules while others are better handled through monitoring, investigation or broader policy controls.

The control-mapping score was calculated over all processed alerts and their top retrieved controls. Confidence scores were generated by the Rank node, but the evaluation also considered whether the mappings were plausible given the alert type and control family. For example, network attack detection commonly relates to System and Information Integrity monitoring, Incident Response handling and System and Communications Protection boundary controls. These mappings are not exact legal conclusions; they are candidate compliance associations that can accelerate analyst review.

Ground truth for Phase 1 was established by matching expected attack scenarios to observed Suricata alerts and registry entries. The evaluation did not rely on subjective model judgment to determine whether an alert existed. Instead, the presence of expected SIDs and corresponding pipeline artifacts was used as deterministic evidence. This is important because LLM-generated explanations may be fluent even when an underlying event is missing; ACE's evaluation therefore separates event presence from explanation quality.

\begin{table}[htbp]

\centering

\caption{Evaluation framework and results} \label{tab:results}

\small

\begin{tabular}{>{\raggedright\arraybackslash}p{0.14\textwidth} >{\raggedright\arraybackslash}p{0.31\textwidth} >{\raggedright\arraybackslash}p{0.47\textwidth}}

\toprule

\textbf{Phase} & \textbf{Metric} & \textbf{Result} \\

\midrule

Phase 1 & Alert processing and NIST mapping & 100\% (8/8 alerts processed and mapped) \\

Phase 2 & False-positive rate & 0\% (0/3 legitimate tests triggered new alerts) \\

Phase 3 & Rule-generation rate & 37.5\% (3/8 alerts produced deployable AUTO rules) \\

Phase 3 & Deployment rate & 100\% (6/6 generated rules confirmed deployed) \\

Phase 4 & Average NIST mapping confidence & 0.87 across all evaluated alerts \\

Phase 5 & Answer relevancy & 1.0, pass \\

Phase 5 & Faithfulness & 1.0, pass \\

Phase 5 & Hallucination metric & 0.67, interpreted as LLM-as-judge calibration issue \\

Performance & Full pipeline runtime & Approximately 173 seconds \\

\bottomrule

\end{tabular}

\end{table}

The evaluation metrics were grouped into five phases. Phase 1 measured true-positive alert processing and NIST mapping. Phase 2 measured false-positive behavior under legitimate traffic. Phase 3 measured rule generation and deployment. Phase 4 measured NIST mapping confidence. Phase 5 measured LLM output quality. 
Table~\ref{tab:results} summarizes the evaluation framework and results.

The methodology emphasizes reproducibility and traceability. Every alert is processed through a fixed sequence of nodes. Each node updates only the state fields assigned to it. The persistent registry records processed SIDs and generated rules. This provides an audit trail suitable for later review by SOC analysts or compliance staff.

\section{Results}

ACE processed all eight attack alerts through the pipeline. This produced a 100\% alert processing and NIST-mapping rate for the evaluated attack set. This result means that each scenario produced a Suricata alert that was ingested, interpreted and mapped to NIST controls. It does not mean that every alert produced a deployable countermeasure rule. The distinction between alert processing and rule generation is important because not every IDS alert contains an application-layer indicator that can be safely converted into a specific rule.

The rule-generation rate was 37.5\%, corresponding to three of eight alerts that produced deployable indicator-specific AUTO rules. Six rules were generated in total. SQL injection produced three rules based on HTTP user-agent, host and URI indicators. Suspicious HTTP user-agent activity produced two rules based on user-agent and host fields. The Nmap SYN scan produced one fallback rule based on source IP. The remaining alerts did not produce deployable indicator-specific rules because they lacked suitable extractable metadata or because the available indicator did not satisfy extraction constraints.

All six generated rules were verified as deployed, resulting in a 100\% deployment success rate for generated rules. This confirms that ACE was able to complete the end-to-end path from alert ingestion through rule proposal to rule placement on the Windows IDS host. Deployment verification is an important result because many AI security prototypes stop at recommendation generation; ACE demonstrates an auditable operational action, while still constraining that action through deterministic templates.

The false-positive evaluation produced no new alerts during the three legitimate traffic tests. This resulted in a measured false-positive rate of 0\% for the specific benign activities tested. The result should be interpreted cautiously because the legitimate test set was small. However, it indicates that ACE's generated rules did not immediately interfere with the tested normal HTTP or DNS activity.

The average NIST control-mapping confidence score across all eight alerts and three retrieved controls per alert was 0.87. This suggests strong semantic alignment between the interpreted threats and retrieved controls. The most relevant control families were those commonly associated with intrusion detection and response, including System and Information Integrity, System and Communications Protection, Incident Response, Risk Assessment and Audit and Accountability. The result supports the use of RAG as a compliance-mapping layer because the system can retrieve and rank controls based on alert semantics rather than relying on brittle keyword mapping.

DeepEval assessment produced scores of 1.0 for answer relevancy and 1.0 for faithfulness, exceeding the evaluation threshold. The hallucination score was 0.67 across two of three cases. This result was not treated as proof that the pipeline hallucinated unsafe content. Instead, it was interpreted in light of known limitations of LLM-as-a-judge evaluation, where different hallucination, groundedness, and faithfulness metrics may diverge on the same output. The deterministic evidence from retrieved controls, rule templates, and deployment verification remains the primary operational validation.

The full eight-alert pipeline run required approximately 173 seconds. This runtime reflects LLM calls, retrieval operations, rule generation, and deployment overhead in a laboratory environment. For SOC use, runtime would need to be optimized through batching, caching, local model options or selective invocation. The SID-based registry already reduces unnecessary repeat processing by skipping previously processed alerts.

Another observation concerns cost and scalability. LLM calls occurred only for alerts that passed deduplication. This is important for operational economics because high-volume IDS environments can generate repeated signatures. By skipping previously processed SIDs, ACE reduces both processing time and model invocation cost. In an enterprise deployment, additional filters could suppress known-benign alerts, prioritize high-severity events or batch similar alerts before invoking interpretation.

The result also highlights the importance of reporting multiple metrics. A single detection metric would make ACE appear uniformly successful, while a single rule-generation metric would understate its compliance-mapping value. The combination of detection, false-positive, deployment, mapping, and LLM-quality measures provides a more complete assessment. For example, a system that generated many rules but produced false positives would be operationally risky. A system that mapped controls accurately but could not deploy rules would be useful for compliance documentation but not for response automation. ACE achieved strong results in detection, deployment, and mapping while showing a clear design boundary in rule-generation coverage.

The per-scenario behavior provides additional insight. The two web-application scenarios produced the richest rule-generation output because their alerts contained HTTP fields. These fields are suitable for content-based Suricata rules. The SYN scan produced a source-IP fallback rule, which is operationally useful in a laboratory but must be handled carefully in production because source IP can be spoofed, shared, or dynamic. The FIN, XMAS, ICMP, and UDP scenarios were successfully mapped to controls but did not produce indicator-specific rules. This pattern supports the claim that ACE is strongest as an integrated interpretation and compliance-mapping system, with rule generation as a conditional capability.

Overall, the results support the feasibility of ACE as an integrated detection-to-compliance-to-response framework. The most robust findings are the successful processing of all attack alerts, absence of false positives in the legitimate tests, successful deployment of all generated rules and high average NIST mapping confidence. The most important limitation is that rule generation was possible only when alerts contained suitable indicators. This limitation is expected and points to a design boundary rather than a system failure.

\section{Discussion}

The results indicate that AI-orchestrated workflows can help close the operational gap between IDS telemetry and compliance documentation. SOC analysts are frequently required to interpret technical alerts and explain their relevance to policy, audit or control frameworks. ACE demonstrates that this translation can be partially automated by combining IDS metadata, LLM-based interpretation and retrieval-grounded control mapping. The 0.87 average confidence score suggests that control ranking can produce useful candidate mappings for analyst review.

The strongest practical value of ACE is not simply faster alert triage. Its value lies in creating a repeatable chain of evidence: raw alert, interpreted threat description, retrieved controls, ranked control mapping, generated rule, and deployment verification. This chain is important for regulated environments because security teams must often demonstrate not only that a threat was detected but also how the detection relates to required controls and what response action was taken.

The results also show why autonomous countermeasure generation must be constrained. ACE generated rules for only three of eight alerts because not all alerts contained indicators suitable for safe rule creation. This conservative behavior is preferable to aggressive rule generation that could block legitimate traffic or create operational instability. For network-layer attacks such as FIN scans, XMAS scans, and UDP floods, ACE currently provides stronger value through compliance mapping and documentation than through indicator-specific rule generation.

The deterministic Supervisor node is a key governance feature. Security workflows that allow LLMs to control routing or tool execution can be difficult to audit. ACE avoids this by using the LLM only where semantic interpretation is required and by keeping routing, deduplication, rule construction and deployment logic deterministic. This design supports explainability, repeatability and safer operational behavior.

The SID registry provides a second governance control. Persistent memory in agentic systems can introduce risk if it is unstructured or contaminated. ACE uses memory in a narrow operational sense: to track processed SIDs and generated rules. This reduces repeated processing and supports idempotent pipeline runs. The registry also gives auditors a record of what the system processed and why repeated deployments were avoided.

From a compliance perspective, ACE is best understood as a decision-support and evidence-generation system, not as a final authority. The control mappings and confidence scores can reduce manual effort, but a production implementation should preserve human review for high-impact deployment decisions. Human-in-the-loop review would be especially important when generated rules could affect business-critical traffic or when mappings are used in formal audit responses.

Finally, ACE illustrates how compliance and response can be designed as mutually reinforcing processes. Compliance controls describe expected security capabilities, while operational alerts provide evidence about whether those capabilities are needed or active. By linking alerts to controls in near real time, organizations can make compliance evidence more operational and make security response more accountable.

The study also suggests a taxonomy of automation outcomes. Some alerts are detection-and-documentation events, where the main output is a threat description and control mapping. Some alerts are detection-documentation-and-rule events, where the alert also contains an actionable indicator. A smaller set may require human escalation because automated rule generation would be unsafe. Distinguishing these categories is more realistic than assuming every alert should trigger an automated control change.

From a risk-management perspective, the framework supports layered assurance. RAG reduces hallucination risk by grounding model output in retrieved documents. Deterministic routing reduces workflow unpredictability. Rule templates reduce syntax and policy risk. Deployment verification confirms whether the intended change occurred. Registry-based deduplication reduces repeated or inconsistent action. No single layer is sufficient on its own, but together they provide a practical pattern for safer AI-enabled security automation.

ACE also has implications for audit readiness. Security teams often prepare evidence after a control assessment has already begun, forcing analysts to search historical logs and reconstruct decisions. ACE's workflow creates evidence during the detection process. The resulting artifacts can show which alert was detected, which controls were retrieved, what confidence scores were assigned and whether a countermeasure was deployed. This does not eliminate the need for audit judgment, but it improves the completeness and timeliness of the evidence trail.

These findings have implications for SIEM and SOC tool design. Future security platforms should not treat compliance as an after-the-fact reporting function. Instead, compliance context can be integrated directly into detection and response workflows. ACE illustrates one way to do this by aligning low-level telemetry with a structured control catalog and carrying the mapping forward into response artifacts.

\section{Limitations and future work}

This study has several limitations. First, the evaluation was conducted in a controlled two-VM environment. The system was not tested against enterprise-scale traffic, multiple network segments or heterogeneous IDS sensors. Second, the attack set was limited to eight scenarios. Although the scenarios cover reconnaissance, web attacks, DNS abuse and denial-of-service activity, they do not represent the full range of modern intrusion behavior. Third, the legitimate traffic set included only three tests, which limits the generalizability of the false-positive result.

Validity threats should also be considered. Internal validity is affected by the controlled nature of the laboratory and by the use of seed rules that may not represent all production detection logic. Construct validity is affected by using confidence scores as a proxy for mapping quality; future work should include expert analyst ratings. External validity is limited by the small number of scenarios and by the two-VM network design. Reliability is supported by deterministic state transitions and registry records, but independent replication would strengthen the findings.

Future studies should include comparative baselines. Useful baselines include manual analyst mapping, keyword-only control retrieval, SIEM-only enrichment and RAG without LLM ranking. Comparing ACE against these alternatives would clarify which parts of the architecture produce the greatest value. Ablation testing could also remove the registry, ranking step or RAG layer to measure their individual contribution.

Fourth, rule generation depends on the presence of extractable indicators. Alerts that lack HTTP, TLS, DNS or payload fields may not produce deployable rules. This is a deliberate safety constraint, but it limits coverage for transport-layer or volumetric events. Fifth, the current RAG implementation uses single-stage retrieval. Incorporating re-ranking, graph-based retrieval or control-family constraints may improve precision. Sixth, LLM output quality was assessed with DeepEval, which provides useful signals but should not be treated as a complete measure of correctness.

Future work should evaluate ACE with larger datasets, additional IDS tools, realistic enterprise traffic and analyst review. The system should be extended with human-in-the-loop approval for high-impact rule deployment, richer rule validation, rollback support and integration with SIEM case-management workflows. Additional research should also compare different LLMs, embedding models and RAG retrieval strategies. A production-oriented version should include authentication hardening, secrets management, encrypted transport, prompt-injection safeguards, access controls and continuous monitoring of generated rule behavior.

Another future direction is to extend ACE beyond NIST SP 800-53 Rev. 5. Many organizations operate under multiple frameworks such as ISO/IEC 27001, CIS Controls, PCI DSS or sector-specific requirements. A multi-framework mapping layer could allow the same alert to generate evidence across several compliance regimes. This would further reduce the manual burden placed on SOC and governance teams.

\section{Conclusion}

This paper presented the Automated Compliance Engine, an AI-orchestrated framework for connecting intrusion detection, NIST SP 800-53 Rev. 5 compliance mapping and automated Suricata rule deployment. ACE integrates Suricata IDS, LangGraph orchestration, RAG-based retrieval, LLM-assisted control ranking, deterministic rule generation, SSH-based deployment and SID-based deduplication.

The laboratory evaluation showed that ACE processed all eight attack alerts, produced no false positives across three legitimate traffic tests, generated deployable rules for 37.5\% of alerts, verified deployment of all generated rules and achieved an average NIST mapping confidence score of 0.87. These findings support the feasibility of an integrated detection-to-compliance-to-response workflow.

ACE should not be interpreted as a complete replacement for SOC analysts or governance professionals. Its primary contribution is to automate repetitive translation and documentation work while preserving auditable evidence and deterministic safeguards. With additional enterprise-scale testing, human-in-the-loop controls and broader framework support, the approach can contribute to more efficient and accountable security operations.


\bibliographystyle{elsarticle-num}
\bibliography{references}

\end{document}